# Controlling Single-Pulse Magnetization Switching through Angular Momentum Reservoir Engineering

B. Kunyangyuen, G. Malinowski, D. Lacour, B. Seng, W. Zhang, S. Mangin, J. Hohlfeld, J. Gorchon, and M. Hehn

*Université de Lorraine, CNRS, IJL, F-54000 Nancy, France.*

We report a systematic study of single-pulse all-optical helicity-independent switching in Co/Gd bilayers, revealing that the magnetization reversal dynamics can be tuned over more than three orders of magnitude. By varying the Gd thickness or inserting a Pt spacer layer between Co and Gd, we control the angular momentum transferred from the rare-earth sublattice to the transition-metal sublattice. Our results show that when Gd is abundant and strongly coupled to Co, angular momentum is efficiently transferred during Gd demagnetization, leading to ultrafast (few-picosecond) Co reversal. In contrast, reducing the Gd thickness or introducing a Pt barrier impedes this transfer, resulting in a domain growth mediated reversal on nanosecond and possibly to microsecond timescales as previously observed in CoDy and CoHo alloys. As a result, in rare-earth–transition-metal systems, replacing Gd with heavier rare-earth elements such as Dy or Ho slows down the switching due to a reduced angular momentum transfer towards Co upon demagnetization as demonstrated in CoGdDy alloys. Our findings establish angular momentum availability and transfer pathways as key parameters governing AO-HIS dynamics, offering a unified framework for fast and slow magnetization reversal across rare-earth–transition-metal systems.

The ability to control magnetization with ultrashort laser pulses has opened promising avenues for next-generation data storage technologies, offering sub-picosecond operation and field-free switching. A key breakthrough was the demonstration of all-optical helicity-independent switching (AO-HIS) in GdFeCo alloys, where a single femtosecond laser pulse reverses magnetization without any applied magnetic field, with switching times on the order of 1 picosecond and excellent energy efficiency [1, 2]. Since then, AO-HIS has been observed in a variety of Gd-based ferrimagnetic alloys [3, 4, 5, 6], in multilayers [7, 8, 9, 10, 11], and in synthetic ferrimagnetic multilayers [8, 12]. The only Gd-free material known to exhibit ultrafast reversal is the ferrimagnet $Mn_2Ru_xGa$ alloy [13]. In Gd-based systems, ultrafast switching is understood to arise from angular momentum transfer generated during the demagnetization of Gd, which acts on the transition metal sublattice when it is also demagnetized.

In the case of Gd-free rare-earth–transition-metal (RE–TM) systems, AO-HIS has been theoretically predicted in TbFe alloys [14], but only partial switching has been observed experimentally [15]. These results demonstrate that, starting from a saturated state, a single laser pulse can reverse the magnetization—i.e., drive it across zero—but the reversal remains incomplete. To the best of our knowledge, no time-resolved magnetization dynamics have been reported in such systems. More recently, full single-pulse all-optical switching has been observed in Co/Ho multilayers as well as in CoHo and CoDy alloys [16, 17]. However, in these cases, deterministic AO-HIS occurs only within a narrow compositional window, and the switching dynamics are extremely slow compared to Gd-based systems. We have shown that magnetization reversal in these systems proceeds via domain coalescence and domain wall reorganization, with characteristic timescales extending up to microseconds. This slow reversal is believed to originate from the gradual evolution of an unbalanced domain configuration, seeded by spin currents generated during the demagnetization of the rare-earth sublattice.

These observations underscore the critical role of angular momentum availability and transfer dynamics in determining the speed and efficiency of AO-HIS. Among rare-earth elements, Gd stands out as an effective angular momentum reservoir, likely due to its relatively long demagnetization time and weak spin–orbit coupling, which facilitate efficient angular momentum transfer to the transition-metal sublattice. In contrast, heavier rare-earth elements such as Tb, Dy, and Ho exhibit strong spin–orbit coupling which significantly reduces angular momentum transfer to Co during demagnetization, thereby limiting their ability to drive ultrafast magnetization reversal [18] [19].

It is worth noting that single-pulse all-optical switching has also been reported in rare-earth–transition-metal (RE–TM) systems such as Co/Tb multilayers and other RE–TM trilayer structures. This type of switching, fundamentally different from AO-HIS, is referred to as all-optical precessional switching (AO-PS) and occurs on the ~100 picosecond timescale [20, 21, 22]. AO-PS is understood to rely on a laser-induced reorientation of magnetic anisotropy—from out-of-plane to in-plane—causing the magnetization to precess around an in-plane axis. This mechanism is thus distinct from the deterministic switching processes described above.

In this context, we investigate how tailoring the rare-earth sublattice—through its composition and interfacial coupling to the transition metal—modulates AO-HIS dynamics. We conduct a systematic study of Co/Gd bilayers by varying the Gd thickness, as well as of Co/Pt/Gd structures, where the insertion of a Pt spacer tunes the efficiency of angular momentum transfer. Our results show that increasing the Gd thickness accelerates switching, while introducing a Pt layer at the Co/Gd interface impedes angular momentum transfer, thereby slowing the reversal process. We further extend our analysis to CoGdDy alloys with constant Co content and demonstrate that increasing the Gd fraction relative to Dy leads to faster switching, reinforcing Gd's role as an efficient angular momentum source. These findings provide a unified framework to understand AO-HIS across RE–TM systems, linking ultrafast switching in Gd-rich compounds to slower, domain-mediated reversal in Dy- and Ho-based alloys, and to partial switching in DyCo and TbCo systems. Overall, our results support the view that the rare-earth element acts as an angular momentum reservoir capable of transferring spin angular momentum to the transition-metal sublattice when it is sufficiently demagnetized. Among these, Gd emerges as the most effective reservoir for enabling fast and deterministic all-optical switching due to its low spin orbit coupling and so efficient transfer of angular momentum to the transition-metal sublattice.

The samples studied in this work were grown by magnetron sputtering on $Si/SiO_2$ (500 nm) or glass substrates, following the deposition of a Ta/Pt seed bilayer. Three types of structures were fabricated: Co(0.7)/Gd($t_{Gd}$), Co(0.7)/Pt($t_{Pt}$)/Gd(3)/Pt(3) wedge multilayers, and $Gd_xDy_{1-x-y}Co_y$ (10 nm) alloys, all capped with a Pt layer to prevent oxidation. The amorphous Ta buffer layer serves two key purposes: it promotes good adhesion of the multilayers to the substrate and helps stabilize the (111) crystallographic texture, particularly important for perpendicular to film plane magnetization in the Pt/Co bilayer. During deposition, the substrate was continuously rotated to ensure uniform film thickness, except during the deposition of the Gd($t_{Gd}$) or Pt($t_{Pt}$) layers, where rotation was halted to induce a controlled thickness gradient. As demonstrated in a previous study [22], this approach enables precise control of layer thickness across the substrate.

Optical switching and time-resolved magneto-optical Kerr effect (TR-MOKE) experiments were performed using a Ti:sapphire femtosecond laser source coupled with a regenerative amplifier. For single-shot optical switching, the experiments were conducted without any applied magnetic field, and the magnetic state was observed several seconds after the laser pulse. In the case of TR-MOKE measurements, a magnetic field was applied perpendicular to the film plane during the experiment to reset the magnetic state between pulses. A delay line of up to 1.6 ns was used to control the temporal delay of the 400 nm probe pulse relative to the 800 nm pump pulse.

As a first step, we investigated the Co(0.7)/Gd($t_{Gd}$) system, specifically Si/SiO(500)/Ta(3)/Pt(3.7)/Co(0.7)/Gd($t_{Gd}$)/Pt(3) multilayers, with Gd thicknesses ranging from 0.15 to 3 nm. This range was selected based on [23], which reported a transition from single-pulse AO-HIS to a multidomain state around $t_{Gd} \approx 0.25$ nm. Polar MOKE measurements (Figure S1a) show square hysteresis loops for all thicknesses, confirming strong perpendicular magnetic anisotropy (PMA) induced by the Pt/Co interface and the (111) crystallographic texture. The coercivity evolves significantly with Gd thickness, reflecting changes in magnetic properties such as reduced saturation magnetization due to antiparallel Co–Gd coupling, modified anisotropy at the top Co/Gd/Pt interface, and altered exchange stiffness resulting from CoGd alloy at the Co/Gd interface.

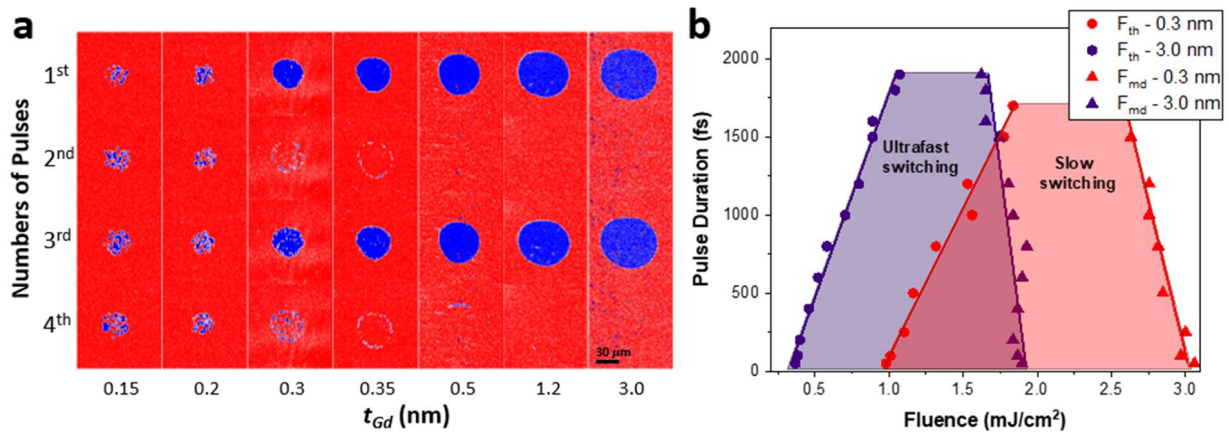

*Figure 1 | AO-HIS state diagram of Co(0.7 nm)/Gd($t_{Gd}$)/Pt structure .* (a) All-optical helicity-independent switching (AO-HIS): Images of the stabilized magnetic states observed after four single laser pulses (fluence: 1.4 mJ/cm², pulse duration: 50 fs) for various Gd thicknesses in Si/SiO(500)/Ta(3)/Pt(3.7)/Co(0.7)/Gd($t_{Gd}$)/Pt(3) wedge samples.(b) AO-HIS threshold fluence ($F_{th}$) and multidomain formation threshold ($F_{md}$) as a function of pulse duration for two representative Gd thicknesses: $t_{Gd}$ = 3 nm (violet) and $t_{Gd}$ )= 0.3 nm (red).

As shown in Figure 1a, clear and reproducible magnetization reversal after four single laser pulses is observed for Gd thicknesses down to 0.3nm. This is consistent with previous findings reported in [23], confirming that AO-HIS persists in Co/Gd bilayers even at very low Gd thickness. The critical fluence required to induce magnetization reversal ($F_{th}$), as well as the fluence threshold for the onset of multidomain formation ($F_{md}$), follow the characteristic triangular shape of AO-HIS state diagrams as a function of pulse duration [6]. Both $F_{th}$ and $F_{md}$ decrease with increasing Gd thickness (Figures 1b and 2b),

The observed decrease in $F_{md}$ with increasing Gd thickness may be linked to the effective Curie temperature of the bilayer—if such a concept can be meaningfully defined in this context. A lower ordering temperature would imply that a smaller laser fluence is sufficient to induce demagnetization. While $F_{th}$ is also influenced by the effective Curie temperature, it could be more directly governed by the spin current generated during Gd demagnetization. According to the criteria outlined in [24], $F_{th}$ scales with the temporal rate of change of the Gd magnetization, i.e., the spin current $j_S$=-d$M_{Gd}$/dt. Thus, increasing the Gd thickness enhances the total angular momentum transferred to the Co sublattice, thereby reducing the fluence required to achieve switching. The energy absorbed by each layer has been quantitatively calculated for samples with Gd thicknesses of 0.3 nm and 3 nm, as detailed in the Supplementary Information S2. The small difference in optical absorption between these two cases is insufficient to account for the significant changes observed in the AO-HIS state diagram (Figure 1b), further emphasizing the dominant role of angular momentum transfer over purely thermal effects.

To gain deeper insight into the switching mechanism, we performed time-resolved magneto-optical Kerr effect (TR-MOKE) measurements (Figure 2a). The reversal dynamics change drastically with the thickness of Gd. For a 3 nm Gd layer, zero magnetization crossing occurs at 1 ps and complete reversal is achieved within 8 ps. As the Gd thickness is reduced, the reversal slows significantly: a demagnetization is observed initially, and the time required to reach full reversal can exceed 1.6 ns—the maximum delay accessible with our setup. Figure 2b provides an estimate of the reversal time as a function of Gd thickness, clearly demonstrating that thinner Gd layers lead to slower switching.

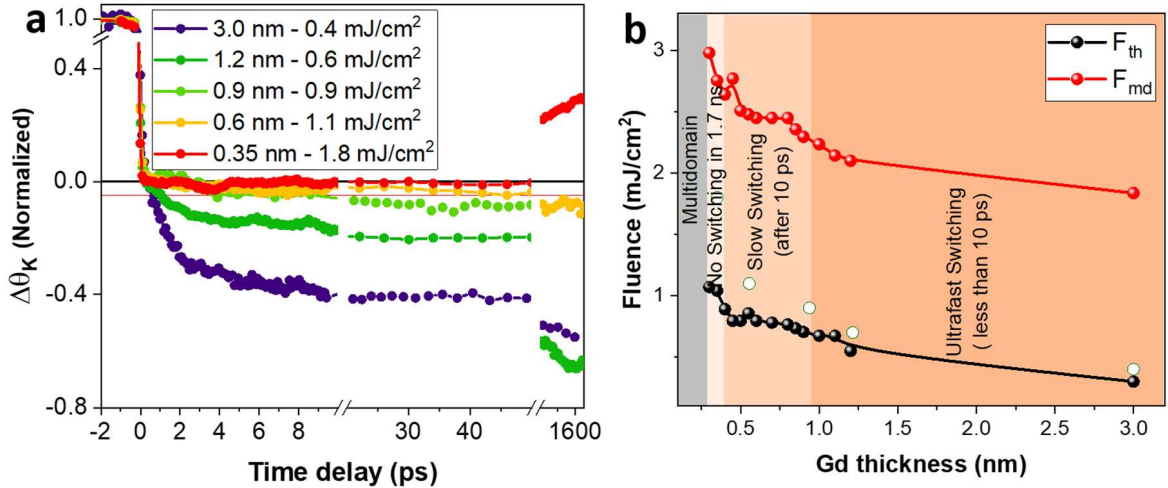

*Figure 2 | Influence of Gd thickness on reversal dynamics and switching thresholds in Co(0.7 nm)/Gd($t_{Gd}$)/Pt.* (a) Time-resolved magneto-optical Kerr effect (TR-MOKE) measurements of magnetization reversal dynamics in Si/SiO(500)/Ta(3)/Pt(3.7)/Co(0.7)/Gd($t_{Gd}$)/Pt(3) wedge samples, performed with a 50 fs laser pulse and an out-of-plane applied magnetic field of 60 mT.
(b) Threshold fluence for all-optical helicity-independent switching ($F_{th}$) and for multidomain formation ($F_{md}$) as a function of Gd thickness, measured with 50 fs laser pulses. The color-coded regions indicate the dynamic response: no switching observed within the measurement window (delay line too short), switching (crossing 5% switching) occurring after 10 ps, or within 10 ps.

The observation of a long-lived demagnetized state, persisting over several nanoseconds, is reminiscent of previous reports on Co/Ho multilayers [16] and even microsecond-scale dynamics observed in CoHo and CoDy single-layer alloys [17]. Micromagnetic simulations have shown that, when the magnetization remains perpendicular to the film plane, an ultrashort spin current can induce an unbalanced Co magnetization distribution. This spin current is believed to originate from the demagnetization of the rare-earth sublattice, which is antiparallel to the Co, ultimately driving the system toward a reversed magnetic state. A similar mechanism may explain the change in behavior observed in Co/Gd bilayers as the Gd thickness is reduced. Thinner Gd layers generate less angular momentum or spin current during demagnetization, leading to a transition from ultrafast (picosecond) to much slower (nanosecond or longer) reversal. When the Gd layer is sufficiently thick, a large angular momentum transfer allows the full reversal of the Co subnetwork in a fast and coherent manner. In contrast, when the Gd angular momentum is limited, only a partial reversal is initially achieved, and full switching requires slow domain nucleation, propagation, or domain wall reorganization.

The magnetic properties of the Co/Gd bilayer—such as saturation magnetization ($M_S$), Curie temperature ($T_C$), and magnetic anisotropy—vary significantly with Gd thickness, as the Co layer is directly coupled to Gd. This makes it difficult to isolate the specific role of each parameter on the magnetization reversal process and its dynamics. To address this, we propose two distinct approaches to independently tune the amount of angular momentum or spin current generated by Gd demagnetization and transferred to Co to induce its reversal.

The first approach involves keeping the Gd thickness fixed at 3 nm and inserting a Pt spacer layer between Co and Gd to reduce the angular momentum transfer across the interface. The second approach is to use GdDyCo alloys in which the Curie temperature is maintained constant by fixing the total rare-earth concentration, while the spin angular momentum transferred to the Co sublattice is varied by adjusting the relative concentrations of Gd and Dy.

Gd has a bulk Curie temperature of 293 K [25], and therefore, in isolation, it is expected to be paramagnetic at room temperature. However, in a Co/Gd bilayer, the interfacial Gd becomes magnetically ordered through exchange coupling with the adjacent Co layer. This induced magnetization decreases progressively with distance from the interface, vanishing toward the bulk of the Gd layer. Introducing an insertion layer between Co and Gd allows a controlled decoupling of this exchange interaction, thereby reducing the net magnetization of the Gd layer as the insertion thickness increases. Direct measurement of the Gd magnetic moment is challenging with standard techniques such as SQUID magnetometry; element specific, X-ray magnetic circular dichroism is required for reliable quantification. Nevertheless, AO-HIS can serve as an indirect probe: the loss of switching can be interpreted as a signature that the Gd moment has dropped below the threshold necessary to trigger magnetization reversal in Co. To explore this, we previously studied all-optical switching in Si/SiO$_2$/Ta/Pt/Co/$x$/Gd/Pt multilayers, where $x$ = Pt, Cu, Ta, or Tb [8]. These materials were selected for their specific interfacial properties: Ta to reduce intermixing, Pt as a non-magnetic material known to polarize at FM interfaces, Cu for its long hot-electron mean free path, and Tb as a rare-earth with strong spin–orbit coupling.

We found that AO-HIS is preserved for insertion layers up to 0.8 nm of Tb, 1 nm of Cu, and 1.7 nm of Pt. This indicates that in all these cases, Gd remains magnetically coupled to Co and its moment remains antiparallel to Co's. In contrast, the use of Ta as an insertion layer resulted in the complete loss of AO-HIS, regardless of thickness. Among the materials tested, Pt was found to be the most suitable for preserving optical switching while allowing controlled tuning of angular momentum transfer, which is why it was chosen in the present study.

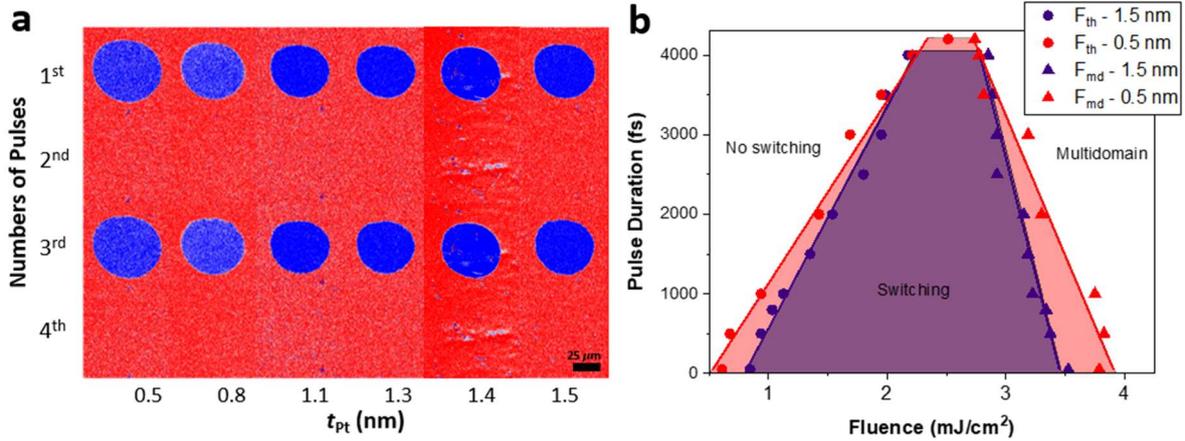

*Figure 3 | AO-HIS state diagram in Co/Pt($t_{Pt}$)/Gd structures using a Pt wedge sample. (a) Magnetic states stabilized after four single laser pulses at a fixed fluence of 1.4 mJ/cm² and pulse duration of 50 fs for different Pt thicknesses on Si/SiO$_2$(500)/Ta(3)/Pt(3.7)/Co(0.7)/Pt(t nm)/Gd(3)/Pt(3) wedge sample. (b) Fluence versus pulse duration state diagrams for Pt spacer thicknesses of 0.5 nm (violet) and 1.47 nm (red), illustrating the characteristic AO-HIS switching window in Co/Pt/Gd structures.*

We studied the structure Si/SiO$_2$(500)/Ta(3)/Pt(3.7)/Co(0.7)/Pt($t_{Pt}$)/Gd(3)/Pt(3), with the Pt spacer thickness varied from 0.5 nm to 1.5 nm. Polar hysteresis loops are presented in Figure S1b and exhibit well-defined square shapes for all Pt thicknesses, confirming robust perpendicular magnetic anisotropy (PMA) promoted by the Pt/Co/Pt interfaces and the (111) crystallographic texture of the Pt/Co stack. The variation in coercivity is less pronounced than in the Co/Gd bilayer series. In first approximation, since alloying between Co and Gd is suppressed by the Pt insertion layer, both the saturation magnetization ($M_s$) and the exchange stiffness of Co can be considered constant. Only slight variations in anisotropy are expected due to the modulation of Pt thickness. Therefore, any observed changes in the magnetization reversal dynamics are less likely to arise from intrinsic magnetic property variations (such as $T_C$), as was the case for direct Co/Gd coupling. As shown in Figure 3a, a robust and reproducible reversal is obtained for all Pt thicknesses using four consecutive pulses with a fluence of 1.4 mJ/cm² and pulse duration of 50 fs. The fluence–pulse duration state diagrams retain their characteristic triangular shape, and both the threshold fluence ($F_{th}$) and the multidomain formation threshold ($F_{md}$) remain mostly independent of the Pt thickness (Figure 3b) in comparison to the Gd thickness influence (Figure 1b). To further elucidate the reversal dynamics, time-resolved magneto-optical Kerr effect (TR-MOKE) measurements were performed and are shown in Figures 4a and 4b.

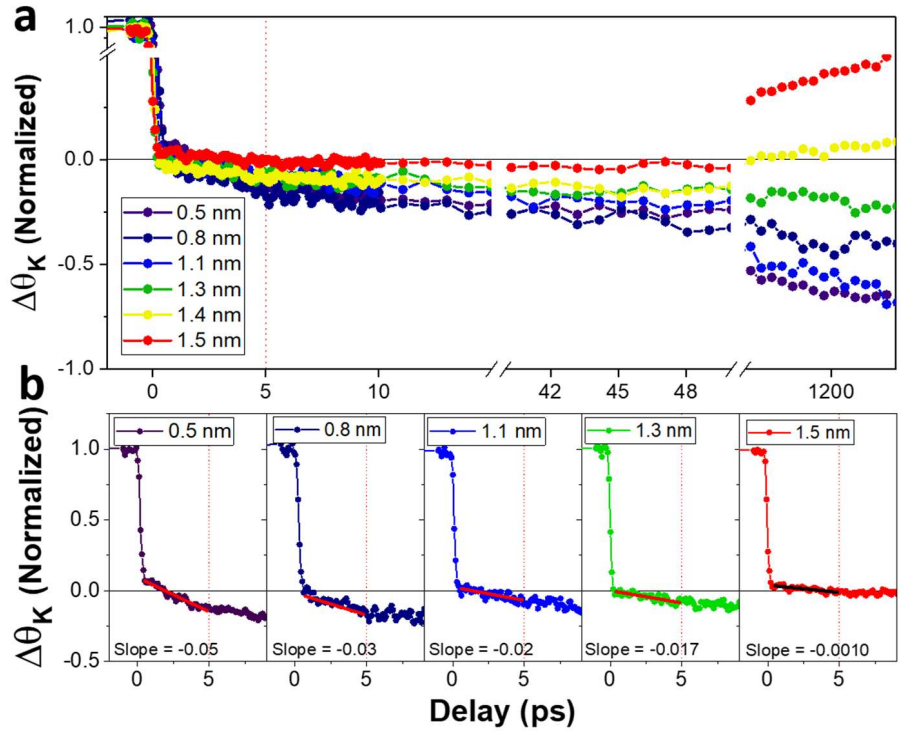

*Figure 4 | Influence of Pt spacer thickness on reversal dynamics in Co/Pt(tPt)/Gd structure. (a) Time-resolved magneto-optical Kerr effect (TR-MOKE) measurements of magnetization reversal dynamics in Si/SiO$_2$(500)/Ta(3)/Pt(3.7)/Co(0.7)/Pt(t)/Gd(3)/Pt(3) wedge samples, for fluences corresponding to the reversal regime observed in static measurements. The measurements were performed with 50 fs laser pulses and an out-of-plane applied magnetic field of 60 mT. (b) Zoom-in on the ultrafast dynamics shown in (a), highlighting the initial demagnetization and reversal onset.*

Once again, varying the Pt insertion layer thickness leads to a clear modification of the magnetization dynamics. For a Pt spacer thickness of 0.5 nm, the reversal process proceeds in three distinct stages: an initial ultrafast demagnetization, followed by a slower reversal driven by angular momentum transfer from the demagnetizing Gd layer, and finally, after approximately 5 ps, relaxation into the reversed magnetic state. As the Pt thickness increases, the overall reversal process becomes progressively slower. The initial demagnetization still occurs, but the subsequent recovery slope diminishes with increasing Pt thickness, delaying the final relaxation into the reversed state. This behavior is consistent with the trends observed in the Co/Gd series and can be attributed to a reduced amount of magnetically coupled Gd. The weakened Co–Gd exchange coupling, along with increased spin scattering in thicker Pt spacers, limits the amount of angular momentum transferred to the Co layer. In the case of the thickest Pt layers, the initial demagnetization remains visible, but the second stage of reversal is barely detectable, and full magnetization switching may take longer than 1 ns to complete.

The second approach to controlling angular momentum transfer involves substituting Dy with Gd in CoDyGd alloys. In CoDy, magnetization reversal is known to be slow [16, 17], as Dy releases significantly less angular momentum during demagnetization compared to Gd. To investigate this effect, we studied a series of $Gd_xDy_{22-x}Co_{78}$ alloys with varying Gd content ($x$ = 6, 10, 13, 16, and 18), while keeping the Co concentration constant. These compositions have been previously shown to support toggle switching behavior [24]. All measurements were conducted at the same laser fluence to isolate the influence of rare-earth composition on the reversal dynamics.

In Figure 5, a clear trend emerges: samples with higher Gd content exhibit significantly faster magnetization reversal compared to those with lower Gd content. In Gd-rich alloys, the transition from demagnetization to full reversal occurs within a few picoseconds, whereas in Dy-rich compositions, it is considerably slower—exceeding 500 ps. This variation cannot be attributed to changes in Curie temperature, as the Co concentration is fixed across the series, and increasing the Gd content slightly raises $T_C$. Taken together with the results from Co/Gd and Co/Pt/Gd systems, these findings confirm the central role of Gd as an efficient angular momentum reservoir. Increasing the Gd fraction enhances the amount of angular momentum transferred to the Co sublattice during laser-induced demagnetization, thereby enabling faster and more efficient all-optical magnetization switching.

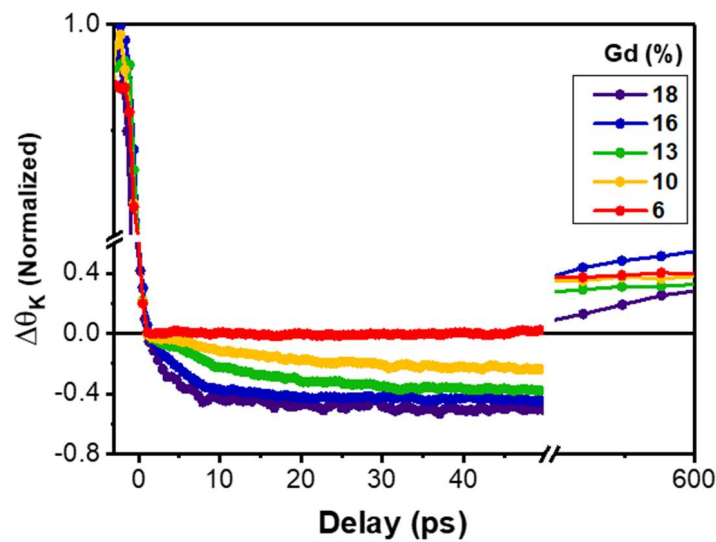

*Figure 5 | **Influence of Gd content on reversal dynamics in CoDyGd alloys.** Time-resolved magneto-optical Kerr effect (TR-MOKE) measurements of magnetization reversal in $Gd_xDy_{22-x}Co_{78}$ alloys with varying Gd content (x = 6, 10, 13, 16, and 18), performed at constant laser fluence. Increasing Gd concentration results in significantly faster reversal dynamics. All samples are taken from Ref. [24].*

These findings provide clear guidance for designing rare-earth–transition-metal (RE–TM) systems optimized for ultrafast and energy-efficient all-optical magnetic switching, particularly by favoring Gd-rich compositions. When the magnetization remains perpendicular to the film and the system passes through a fully demagnetized state, a spin current is generated by the laser-induced demagnetization of the RE sublattice. If the RE magnetization is antiparallel to the Co subnetwork, this spin current induces an unbalanced magnetization distribution in the Co, which subsequently evolves into a reversed magnetic state. This process typically occurs on the microsecond timescale and involves the transfer of a small portion of the RE's spin angular momentum to the Co sublattice, while the majority is dissipated into the lattice due to the strong spin–orbit coupling of the RE element. A similar behavior is observed in Co/Gd bilayers with very thin Gd layers, where the limited angular momentum transfer to Co results in slower, domain-mediated switching.

Taken together, our results establish a comprehensive framework for understanding and controlling all-optical switching across a wide class of RE–TM systems. By systematically tuning rare-earth composition and interfacial coupling, we have demonstrated how the efficiency and speed of magnetization reversal are governed by angular momentum availability and transfer dynamics. This unified picture links ultrafast switching in Gd-rich systems to slower, domain-driven processes in Dy- and Ho-based alloys, and highlights the critical role of interface engineering in tailoring spin current generation. These insights pave the way for the rational design of RE–TM materials for next-generation opto-spintronic devices.

Acknowledgement

This work is supported by the ANR-17-CE24-0007 UFO project and ANR-23-CE30-0047 SLAM project, the Institute Carnot ICEEL, the Région Grand Est, the Metropole Grand Nancy for the project "OPTIMAG" and FASTNESS, the interdisciplinary project LUE "MAT-PULSE", part of the French PIA project "Lorraine Université d'Excellence" reference ANR-15-IDEX-04-LUE. This work was supported by the ANR through the France 2030 government grants EMCOM (ANR-22-PEEL-0009), PEPR SPIN – SPINMAT ANR-22-EXSP- 0007, PEPR SPIN (ANR-22-EXSP 0002). This article is based upon work from COST Action CA23136 CHIROMAG. This work also was supported by the NSRF (Thailand) via the Program Management Unit for Human Resources & Institutional Development, Research and Innovation [grant number B05F650024].

Author contributions

G. M., M. H., J. G., and S. M. conceived the research topic. M. H. performed sample fabrication. B. K., B. S. and D. L. performed sample magnetic characterization. J. G., J. H. and

G. M., developed AOS measurement set-up. B. K., W. Z. and G. M. performed AOS experiment and data analysis. M. H., S. M. and G. M. wrote manuscript. All authors contributed to discuss the measurement results.

References

[1] Radu, I.; Vahaplar, K.; Stamm, C.; Kachel, T.; Pontius, N.; Dürr, H. A.; Ostler, T. A.; Barker, J.; Evans, R. F. L.; Chantrell, R. W.; Tsukamoto, A.; Itoh, A.; Kirilyuk, A.; Rasing, Th.; Kimel, A. V. Transient ferromagnetic-like state mediating ultrafast reversal of antiferromagnetically coupled spins. *Nature* **2011,** 472, 205.

[2] Ostler, T.A.; Barker, J.; Evans, R.F.L.; Chantrell, R.W.; Atxitia, U.; Chubykalo-Fesenko, O.; El Moussaoui, S.; Le Guyader, L.; Mengotti, E.; Heyderman, L.J.; Nolting, F.; Tsukamoto, A.; Itoh, A.; Afanasiev, D.; Ivanov, B.A.; Kalashnikova, A.M.; Vahaplar, K.; Mentink, J.; Kirilyuk, A.; Rasing, Th.; Kimel, A.V. Ultrafast heating as a sufficient stimulus for magnetization reversal in a ferrimagnet. *Nat. Commun*. 2012, 3, 666.

[3] Wilson, R. B.; Gorchon, J.; Yang, Y.; Lambert, C. H.; Salahuddin, S.; Bokor, J. Ultrafast magnetic switching of GdFeCo with electronic heat currents. *Phys. Rev. B* **2017**, 95, 180409(R).

[4] Xu, Y.; Deb, M.; Malinowski, G.; Hehn, M.; Zhao, W.; Mangin, S. Ultrafast Magnetization Manipulation Using Single Femtosecond Light and Hot-Electron Pulses. *Advanced Materials* **2017**, 29, 1703474.

[5] Yang, Y.; Wilson, R. B.; Gorchon, J.; Lambert, C. H.; Salahuddin, S.; Bokor, J. Ultrafast magnetization reversal by picosecond electrical pulses, *Sci. Adv.* **2017***,* 3, e1603117.

[6] Wei, J.; Zhang, B.; Hehn, M.; Zhang, W.; Malinowski, G.; Xu, Y.; Zhao, W.; Mangin, S. All-optical Helicity-Independent Switching State Diagram in Gd-Fe-Co Alloys, *Phys. Rev. Applied* **2021**, 15, 054065.

[7] Lalieu, M. L.; Peeters, M. J. G.; Haenen, S. R. R.; Lavrijsen, R.; Koopmans, B. Deterministic all-optical switching of synthetic ferrimagnets using single femtosecond laser pulses, *Phys. Rev. B* **2017**, 96, 220411.

[8] B. Seng, Master thesis, Lorraine University (2018).

[9] Li, P.; Peeters, M. J. G.; van Hees, Y. L. W.; Lavrijsen, R.; Koopmans, B. Ultra-low energy threshold engineering for all-optical switching of magnetization in dielectric-coated Co/Gd based synthetic-ferrimagnet, *Appl. Phys. Lett.* **2021**, 119, 252402.

[10] Li, P.; Kools, T. J.; Koopmans, B.; Lavrijsen, R. Ultrafast racetrack based on compensated Co/Gd-based synthetic ferrimagnet with all-optical switching, *Adv. Electron. Mater*. **2022**, 8, 2200613.

[11] Li, P.; Kools, T. J.; Pezeshki, H.; Joosten, J. M. B. E.; Li, J.; Igarashi, J.; Hohlfeld, J.; Lavrijsen, R.; Mangin S.; Malinowski, G.; Koopmans, B. Picosecond all-optical switching of Co/Gd–based synthetic ferrimagnets, *Phys. Rev. B* **2025**, 111, 064421.

[12] Beens, M.; Lalieu, M. L. M.; Deenen, A. J. M.; Duine, R. A.; Koopmans, B. Comparing all-optical switching in synthetic-ferrimagnetic multilayers and alloys, *Phys. Rev. B* **2019**, 100, 220409(R).

[13] Banerjee, C.; Teichert, N.; Siewierska, K. E.; Gercsi, Z.; Atcheson, G. Y. P.; Stamenov, P.; Rode, K.; Coey, J. M. D.; Besbas, J. Single pulse all-optical toggle switching of magnetization without gadolinium in the ferrimagnet $Mn_2Ru_xGa$, *Nature Commun.* **2020**, 11, 4444.

[14] Jiang, C.; Liu, D.; Song, X.; Wu, Y.; Li, H.; Xu, C. Single-shot all-optical switching of magnetization in TbFe, *J. Phys. D: Appl. Phys*. **2024**, 57, 195001.

[15] Hu, Z.; Besbas, J.; Smith, R.; Teichert, N.; Atcheson, G.; Rode, K.; Stamenov, P.; Coey, J. M. D. Single-pulse all-optical partial switching in amorphous $Dy_xCo_{1-x}$ and $Tb_xCo_{1-x}$ with random anisotropy; *Appl. Phys. Lett.* **2022**, 120, 112401.

[16] Peng, Y.; Malinowski, G.; Gorchon, J.; Hohlfeld, J.; Salomoni, D.; Buda-Prejbeanu, L. D.; Sousa, R. C.; Prejbeanu, I. L.; Lacour, D.; Mangin S.; Hehn, M. Observation of Single-shot Helicity-independent all-optical switching in Co/Ho multilayers, *Phys. Rev. Appl*. **2023**, 20, 014068.

[17] Kunyangyuen, B.; Malinowski, G.; Lacour, D.; Lin, J. -X.; Le Guen, Y.; Buda-Prejbeanu, L. D.; Mangin, S.; Gorchon, J.; Hohlfeld, J.; Hehn, M. Single-laser pulse toggle switching in CoHo and CoDy single layer alloys : when domain wall motion matters, arXiv **2025**, 2503.22579.

[18] Zhang, W.; Lin, J. X.; Huang, T. X.; Malinowski, G.; Hehn, M.; Xu, Y.; Mangin, S.; Zhao, W. S. Role of spin-lattice coupling in ultrafast demagnetization and all optical helicity-independent single-shot switching in $Gd_{1-x-y}Tb_yCo_x$ alloys *Phys. Rev. B* **2022**, 105, 054410.

[19] Frietsch, B.; Donges A.; Carley, R.; Teichmann, M.; Bowlan, J.; Döbrich, K.; Carva, K.; Legut, D.; Oppeneer, P. M.; Nowak, U.; Weinelt, M. The role of ultrafast magnon generation in the magnetization dynamics of rare-earth metals, *Sci Adv* **2020**, 6, 39.

[20] Salomoni, D.; Peng, Y.; Farcis, L.; Auffret, S.; Hehn, M.; Malinowski, G.; Mangin, S.; Dieny, B.; Buda-Prejbeanu, L. D.; Sousa, R.C.; Prejbeanu, I. L. Field-free all-optical switching and electrical read-out of Tb/Co-based magnetic tunnel junctions, *Phys. Rev. Appl*. **2023**, 20, 034070.

[21] Mishra, K.; Blank , T. G. H.; Davies, C. S.; Prejbeanu, I. L.; Avilés-Félix, L.; Koopmans, B.; Salomoni, D.; Buda-Prejbeanu, L. D.; Sousa, R. C.; Rasing, Th.; Kimel, A. V.; Kirilyuk, A. Dynamics of all-optical single-shot switching of magnetization in Tb/Co multilayers, *Phys. Rev. Research* **2023**, 5, 023163.

[22] Peng, Y.; Salomoni, D.; Malinowski, G.; Zhang, W.; Hohlfeld, J.; Buda-Prejbeanu, L. D.; Gorchon, J.; Vergès, M.; Lin, J. X.; Sousa, R.C.; Prejbeanu, I. L.; Mangin, S.; Hehn, M. In plane reorientation induced single laser pulse magnetization reversal, *Nature Comm.* **2023**, 14, 5000.

[23] Petty Gweha Nyoma, D.; Hehn, M.; Malinowski, G.; Hohlfeld, J.; Gorchon, J.; Vergès, M.; Mangin, S.; Montaigne F. Gd doping at the Co/Pt interfaces to induce ultra-fast all-optical switching, *ACS Appl. Electron. Mater*. **2024**, 6, 1122.

[24] Zhang, W.; Hohlfeld, J.; Huang, T. X.; Lin, J. X.; Hehn, M.; Le Guen, Y.; Compton-Stewart, J.; Malinowski, G.; Zhao, W. S.; Mangin, S. Criteria to observe single-shot all-optical switching in Gd-based ferrimagnetic alloys, *Phys. Rev. B* **2024**, 109, 094412.

[25] Weller, D.; Alvarado, S. F.; Campagna, M. Bulk and surface curie temperatures of Gd(0001), *Physica B+C* **1985**, 130, 72.

# Supplementary figures

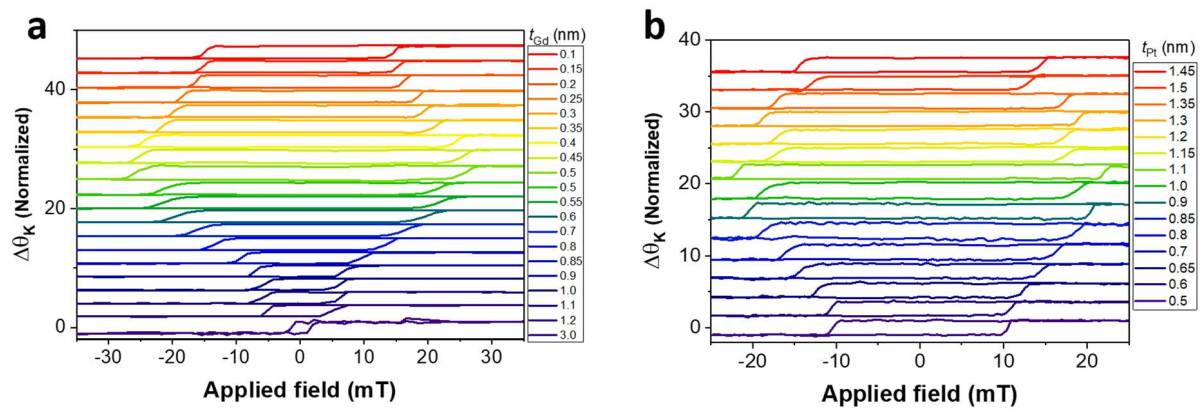

***Figure S1 | Polar MOKE measurements.*** *a) Hysteresis loops of the Si/SiO(500)/Ta(3)/Pt(3.7)/Co(0.7)/Gd($t_{Gd}$)/Pt(3) wedge sample show perpendicular magnetic anisotropy (PMA) in all the thickness, however, the magnetic properties may change as with thicker Gd. b) Hysteresis loops of the Si/SiO$_2$(500)/Ta(3)/Pt(3.7)/Co(0.7)/Pt($t_{Pt}$)/Gd(3)/Pt(3) wedge sample show almost constant of the magnetic properties as a function of Pt thickness.*

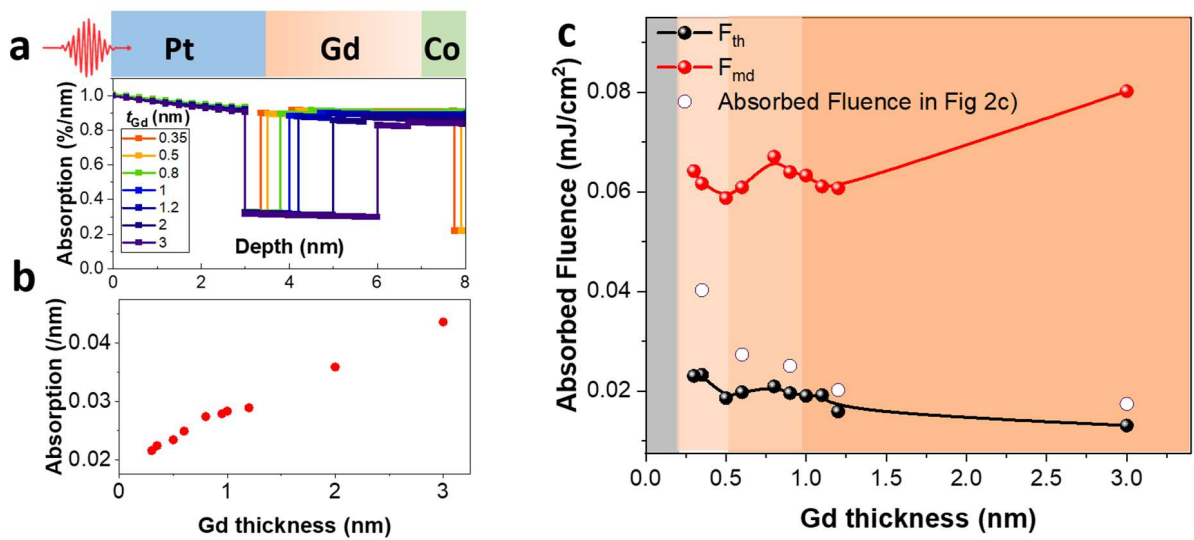

***Figure S2 | Energy absorption profile of Si/SiO(500)/Ta(3)/Pt(3.7)/Co(0.7)/Gd($t_{Gd}$)/Pt(3) samples with wedge of Gd thickness.*** *a) Calculated optical absorption as a function of penetration depth b) Total absorption energy inside the magnetic layers as function of Gd thickness c) The fluences extracted from Figure 2b were normalized by the absorbed energy in the magnetic Co/Gd layers, revealing that the fluence required for switching remains nearly constant across different Gd thicknesses.*